\documentclass[reprint,amsmath,amssymb,aps,prb, superscriptaddress]{revtex4-2}

\usepackage{graphicx}
\usepackage{dcolumn}
\usepackage{bm}
\usepackage{xcolor}

\begin{document}


\title{Interface engineering of spin triplet Cooper pairs for spin-valve implementation}

\author{Debashree Nayak}
\affiliation{School of Physical Sciences, National Institute of Science Education and Research (NISER) Bhubaneswar, Jatni, Odisha-752050, India}
\affiliation{Homi Bhabha National Institute, Training School Complex, Anushakti Nagar, Mumbai-400094, India}
\author{Abhisek Sahoo}
\affiliation{School of Physical Sciences, National Institute of Science Education and Research (NISER) Bhubaneswar, Jatni, Odisha-752050, India}
\affiliation{Homi Bhabha National Institute, Training School Complex, Anushakti Nagar, Mumbai-400094, India}
\author{Pratap Kumar Sahoo}
\affiliation{School of Physical Sciences, National Institute of Science Education and Research (NISER) Bhubaneswar, Jatni, Odisha-752050, India}
\affiliation{Homi Bhabha National Institute, Training School Complex, Anushakti Nagar, Mumbai-400094, India}
\author{Kartik Senapati}%
\email{kartik@niser.ac.in}
\affiliation{School of Physical Sciences, National Institute of Science Education and Research (NISER) Bhubaneswar, Jatni, Odisha-752050, India}
\affiliation{Homi Bhabha National Institute, Training School Complex, Anushakti Nagar, Mumbai-400094, India}

\date{\today}

\begin{abstract}
Spin singlet Cooper pairs can be converted into equal-spin triplet pairs at a superconductor/heavy-metal interface under suitable conditions. Under current carrying conditions, these triplet correlations can give rise to a non equilibrium spin moment in the heavy metal through the predicted supercurrent spin-Hall effect. Here we demonstrate that this current induced triplet spin moment, in conjunction with the magnetic moment of a ferromagnetic layer, can enable a magnetic spin-valve response in Nb/Pt/Ni/Pt/Nb vertical nano-devices. The magnitude of this spin-valve effect depends on the efficiency of singlet-triplet Cooper pair conversion at the Nb/Pt interface. In order to facilitate efficient triplet generation, we deliberately introduced interfacial roughness to introduce finite Rashba spin-orbit coupling at the Nb/Pt interface and an out-of-plane component of magnetic moment at the Ni interface. In contrast, no discernible spin-valve response was observed in devices with smooth interfaces within measurement resolutions.  Since the magnetic moment of the Ni layer and the current-induced triplet spin moment in the Pt layer are independently switchable using a magnetic field and bias current, respectively, the spin valve can be controlled via either parameters. These results demonstrates a novel approach to directly utilizing the spin polarized triplet Copper pairs in superconducting spintronic applications.


\end{abstract}

\maketitle


\section*{\label{sec:level1}Introduction}

Magnetic spin valves are multilayer structures where the electrical resistance is determined by the relative orientation between the individual ferromagnetic layers. Usually the constituent magnetic layers are chosen to have slightly different coercive fields which allows to switch the magnetization of the soft magnetic layer by an external field without rotating the magnetization in the other layer. The resulting parallel (P) and antiparallel (AP) magnetic configurations lead to different values resistances due to spin dependent scattering rates of conduction electrons\cite{baibich,Dieny}. Over the past few decades several material combinations have been explored to realize spin active valve like response, including ferromagnet/non-magnet hetero-structures\cite{F/Nspinvalvejedema,F/Nspinvalvejedema2001,F/Nspinvalvejogo2007}, semiconductor-based systems\cite{Wolf,spinvalvewong2016spin,spinvalvehonda2020germanium},and also, low-dimensional systems\cite{Ghising,Ahn2020}. Heavy metal layers with strong spin-orbit coupling add further tunability to the spin transport in magnetic spin valves devices. For example, spin-current generated via spin Hall effect in a heavy metal layer\cite{Song, Brataas}, such as Pt, can exert a torque in an adjacent magnetic layer. This allows for a charge current based control of the spin valve devices rather than a magnetic field based control.

Incorporating a superconducting layer in the vicinity of magnetic spin-valves has resulted in  spin-valve architectures\cite{buzdin1999,Gu,supspinvalvenowak} where the relative orientation of the magnetic layers controls the transition temperature T$_C$. In these cases a parallel magnetic configuration was shown to have a lower T$_C$ compared to the antiparallel configuration\cite{leksin2015,Robinson}. However, unlike SOC materials, superconductors cannot intrinsically generate a spin-polarization usable in a spin valve structure. This is fundamentally due to the net spin-zero pairing of spin-up ($\uparrow$) and spin-down ($\downarrow$) electrons, in a spin-singlet condensates. However, the equal spin components ($\uparrow\uparrow$ and $\downarrow\downarrow$) of spin-triplet pair function are spin polarized in nature. Therefore, a proximity induced spin-triplet correlation in a non-magnetic layer can develop a polarization, if the $\uparrow\uparrow$ and $\downarrow\downarrow$ components are spatially separated. Theoretically it has been shown that, a non-equilibrium spatial separation of the $\uparrow\uparrow$ and $\downarrow\downarrow$ components is indeed expected in the current carrying state of Josephson junctions \cite{SpinHall_JJ} Though the spatial separation of the $\uparrow\uparrow$ and $\downarrow\downarrow$ components is analogous to the usual spin-Hall effect, we must mention that, no spin-current is expected in this case \cite{SpinHall_JJ}. 

The primary recipe for the generation of odd-frequency, long range triplet correlation has been to pass singlet Cooper pairs through magnetically non-collinear regions \cite{S/Ftripletbergeret,S/Ftripletbergeret1,S/Ftripletkhaire,S/Ftripletvolkov,Tamura_triplet}. The combined effect of spin mixing and spin rotation in a non-collinear magnetic region leads to this conversion. However, alternate routes have also been proposed to realize long range spin-triplet correlations via the interaction of magnetic exchange field with spin-orbit coupling effects, even in the absence of magnetic non-collinearity \cite{eskilt_Linder,bergeret_Tokatly_2013,jacobsen_linder,bergeret2014}. Bergeret \textit{et al.} predicted that a ferromagnetic Josephson junction with Rashba and Dresselhaus spin orbit coupling can support an equal spin triplet supercurrent, provided that the magnetization of the ferromagnetic (F) layer possesses an in-plane component\cite{bergeret2014}. Subsequently, Satchell \textit{et al.} considered the case of only Rashba SOC and showed that both in-plane and out of plane components of magnetization are required\cite{satchell2019supercurrent}.

In this work we show that in a suitable device geometry, the non-magnetic layer with induced spin-triplet polarization can establish a significant spin-valve like coupling with an adjacent magnetic layer. Our device geometry consisted of a stack of Nb-Pt-Ni-Pt-Nb, lithographically patterned into vertical nano-junctions. In the superconducting state of the Nb electrodes, a triplet polarization was induced in the Pt layer, which, in the presence of a bias current, can lead to a triplet spin-Hall effect\cite{SpinHall_JJ}. It is important to note that in the Nb/Pt/Ni/Pt/Nb hetero-structure only the proximity of superconducting Nb is required to generate a triplet correlation in Pt. Superconducting (Josephson) coupling across the entire stack is not essential. The presence of such triplet pairing in superconductors with spin–orbit coupling (SOC) was theoretically predicted by Gor’kov et al.\cite{SSOCtriplettheory}, while its manifestation at the Nb/Pt interface has been experimentally proved through Muon spin resonance ($\mu$SR) measurements \cite{Stephenlee_muSR}, through spin pumping experiment \cite{jeon_spin_pumping}, and more recently via magnetization measurements\cite{NetoTriplet,S/SOCtriplet}. In our device geometry the direction of the induced triplet spin magnetization in Pt does not depend on the magnetization direction of the ferromagnet, rather depends it on the direction of the applied bias current \cite{SDEPtJJ}. On the other hand, for a fixed direction of bias current, the direction of magnetization in Ni can be switched with an external magnetic field. Therefore, effectively we can obtain a spin-valve structure controllable either via bias current or via external magnetic field.  A small change in resistance was indeed observed in our earlier work on Nb/Pt/Ni/Nb junctions \cite{SDEPtJJ} corresponding to a spin-valve like behavior. Here we demonstrate that slight increase in the interfacial roughness in Nb(160nm)/Pt(15nm)/Ni(20nm)/Pt(15nm)/Nb(160nm) vertical device leads to an order of magnitude enhancement in the resistance difference between the parallel and antiparallel states, which implies a much pronounced spin-valve effect. 

\section*{results}
\subsection*{Singlet-Triplet conversion at Nb-Pt interface}
\begin{figure}[t]
    \centering
    \includegraphics[width=\columnwidth]{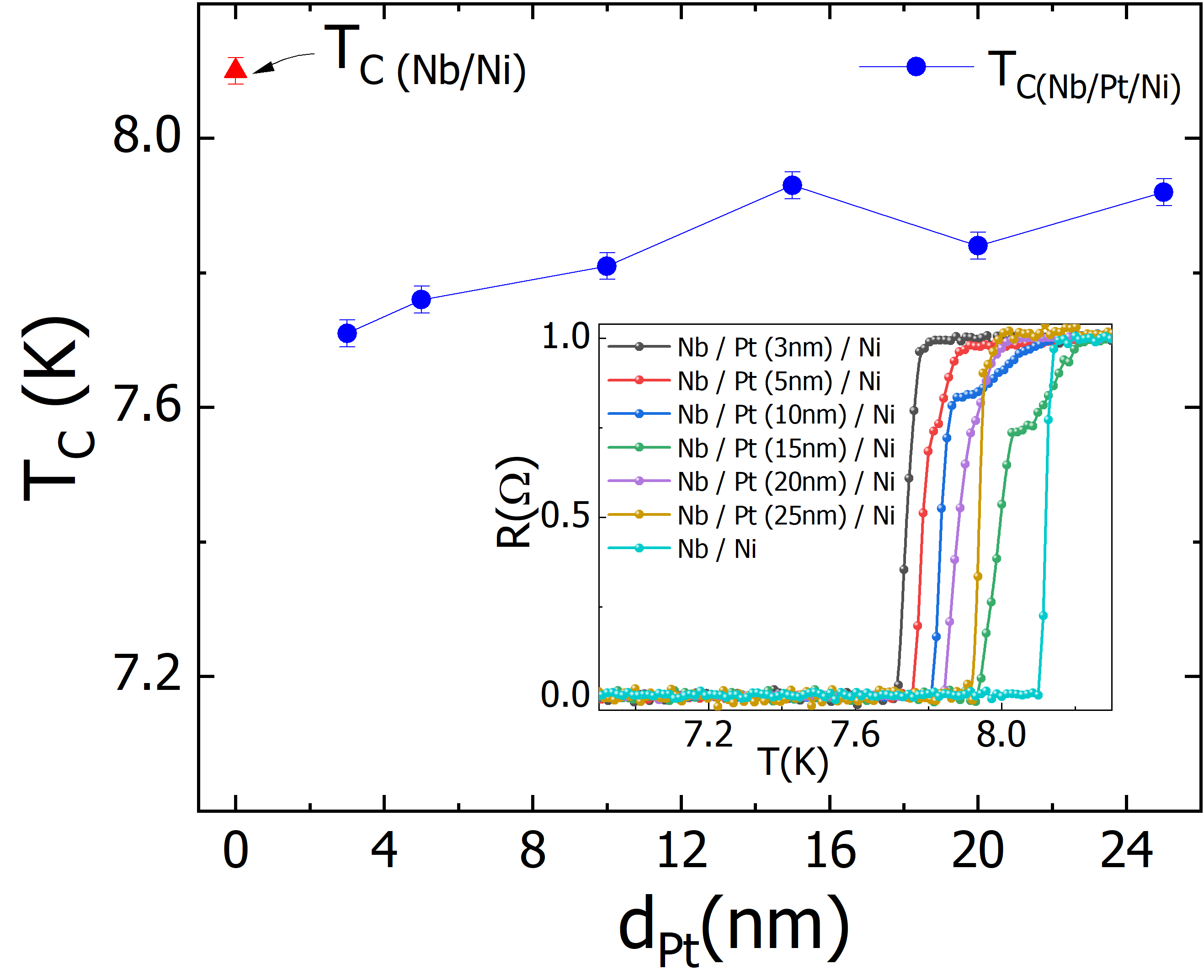} 
    \caption{Superconducting transition temperatures of Nb/Pt/Ni trilayer films are plotted as a function of the Pt layer thickness. The temperature corresponding to zero resistance was taken as the transition temperature. In comparison to the T$_C$ of Nb/Ni bilayer, shown as a triangular symbol in this figure, all Nb/Pt/Ni trilayers showed significant suppression, due to to triplet induced drainage of Cooper pairs from the the Nb layer. The inset shows the R(T) curves for all the trilayers from which the T$_C$ values were extracted. }
    
    \label{Fig.1}
\end{figure}
In the Nb/Pt/Ni/Pt/Nb multilayer stack the simultaneous presence of a ferromagnetic exchange field in Ni and interfacial spin-orbit field at the Nb/Pt interfaces provide the requisite conditions prescribed by Bergeret et al.\cite{bergeret_Tokatly_2013, bergeret2014} for singlet-triplet conversion. Since our proposed spin-valve structure relies on the spin polarization brought in by the equal-spin triplet components, we performed T$_c$ measurements on a series of Nb(40nm)/$\mathrm{Pt}(d_{\textit{Pt}})$/Ni(100nm) trilayer structure to look for signatures of triplet conversion. Earlier studies\cite{Flokstra_suppression,SinghTripletDrainage} have shown that formation of triplet correlations in superconducting hetero-structures leads to a drainage of spin-singlet Cooper pairs leading to a suppression of transition temperature. In the Figure 2 we have plotted the transition temperatures (corresponding to zero resistance) of a series of Nb/Pt/Ni trilayers with varying thickness of Pt layer. The corresponding R(T) curves are plotted in the inset. Compared to the case of a Nb/Ni bilayer we observe a significant suppression in T$_c$ in all Nb/Pt/Ni trilayers. This observation is consistent with the earlier results\cite{Flokstra_suppression, SinghTripletDrainage} and confirms the formation of triplet correlation in the Pt layer. The roughly increasing trend of T$_c$ as a function of Pt thickness indicates that increasing Pt thickness gradually reduces the effect of magnetic exchange field from the Ni, which is an essential ingredient for the singlet-triplet conversion\cite{bergeret_Tokatly_2013, bergeret2014}.  

\subsection*{Interface roughness tuning}
\begin{figure}[t]
    \centering
    \includegraphics[width=\columnwidth]{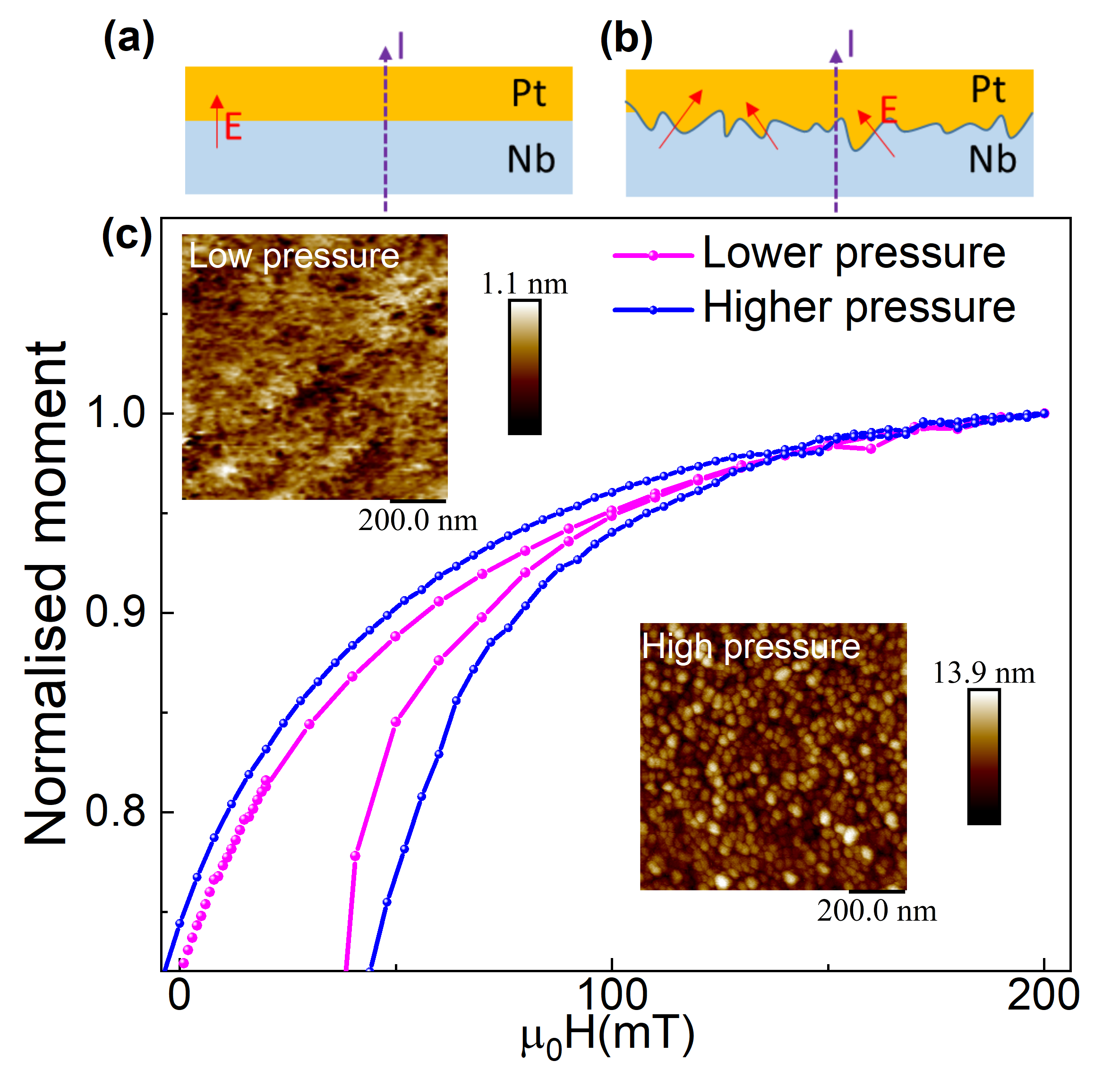} 
    \caption{Panels (a) and (b) compare the schematic diagrams of the Nb/Pt interfaces in the smooth and rough conditions. For the ideally smooth interface the interface Rashba Hamiltonian $H_{R} = \alpha_{R} \, (\boldsymbol{E} \times \mathbf{p}) \cdot \boldsymbol{\sigma}$ vanishes due the zero cross product $\boldsymbol{E} \times \mathbf{p}$. In the presence of a finite roughness at the interface, the $\boldsymbol{E} \times \mathbf{p}$ product need not vanish entirely, leading to a finite interfacial Rashba spin-orbit coupling. (c) Comparison of the equilibrium magnetization of the smooth and rough Nb/Pt/Ni/Pt/Nb multilayers in the first quadrant. The saturation field as well as the coercive field of the rough multilayer were found to be higher than the smooth multilayer. Inset shows the AFM surface topography of the as deposited single layer Nb films under two different pressure conditions, to directly access the roughness which propagates to the further layers deposited on the base Nb. RMS roughness of the film deposited at higher pressure is much higher than the film deposited at lower pressure. }

    \label{Fig.2}
\end{figure}
The other important consideration for the singlet-triplet conversion in our Nb/Pt/Ni/Pt/Nb devices is the presence of a finite interfacial spin-orbit field. For a bias current perpendicular to an ideally flat Nb-Pt interface the Rashba Hamiltonian, given by $H_{R} = \alpha_{R} \, (\boldsymbol{E} \times \mathbf{p}) \cdot \boldsymbol{\sigma}$, becomes zero because the inversion symmetry breaking field $\boldsymbol{E}$ and the electron momentum $\mathbf{p}$ become parallel. Here $\alpha_{R}$ is the Rashba coupling constant and $\boldsymbol{\sigma}$ represents the Pauli spin matrices. However, in real samples the inversion symmetry breaking field need not be perfectly perpendicular to the interface, which allows for a small interfacial Rashba SOC, though the centrosymmetric crystal structure of Pt excludes Dresselhause type SOC in the bulk of Pt. Therefore, we introduce roughness at the interface to generate interfacial electric field components $\boldsymbol{E}$ that are not parallel to the current direction as shown in the schematic in the Fig 2(a). This can contribute to the enhancement of Rashba spin orbit coupling (SOC). It has also been reported that interfacial strain can lead to orbital deformation, which serves as an additional source of Rashba SOC \cite{leeinterfacial_Rashba}.

Interface roughness in our Nb/Pt/Ni/Pt/Nb multilayer stack was introduced by varying the Ar pressure during the deposition of the bottom Nb layer. It is known that increasing sputtering pressure typically increases the surface roughness of magnetron sputtered Nb films, if other deposition parameters are kept constant\cite{RAO_roughness}. Surface roughness of the bottom Nb layer translates to the other layers deposited on it. In order to elucidate the effect of interface, we prepared the bottom Nb layer at two different Argon pressure conditions, keeping the Ar flow rate of 20~SCCM and sputtering power of 55~W fixed. The surface roughness of the 80 nm thick Nb films deposited at high and low Ar pressures were characterized by atomic force microscopy, as shown in the inset of Fig.2(a). At a deposition pressure of $1 \times 10^{-2}$~mbar Nb film exhibited a RMS roughness of approximately 2.1~nm, while a RMS roughness of $\sim$1nm was observed for the film deposited at an order of magnitude lower pressure. Though the bottom layer of the Nb/Pt/Ni/Pt/Nb multilayer stack was deposited at these two deposition pressures, we used equilibrium magnetization measurements of the complete stacks as an indirect means of assessing the differences in interface roughness between the two cases. This is because roughness is expected to influence the saturation behavior of the magnetic Ni layer in the stack through its effect on demagnetization field and magnetic anisotropy\cite{mag_roughness}. The normalized magnetization values at 10 K for the two stacks, in the first quadrant of the MH curves, are compared in the main panel of Fig 2. The samples were mounted in the in-plane applied magnetic field configuration during the MH measurements in a SQUID magnetometer. The hysteresis loops for both samples confirm the in-plane magnetic anisotropy of the 20~nm thick Ni layer. However, the multilayer deposited at a higher pressure exhibited a higher saturation field of approximately 130~mT, compared to the sample deposited at lower pressure, which showed a saturation field of around 90~mT. This indicates additional out-of-plane component of magnetization at the rougher interface, consistent with earlier reports\cite{mag_roughness,mag_roughness2}.

\subsection*{Spin valve response and the effect of roughness engineering}
Vertical nano-pillar devices of typical dimension 300nm $\times$200nm were fabricated from the Nb-Pt-Ni-Pt-Nb multilayer stacks using a combination of optical lithography and focused ion beam milling. High resolution scanning electron micrograph of two representative devices are as shown in Fig 3(a) and (c). Note that the device in Fig 3(a) has a visibly higher interface roughness compared to the Fig 3(c). Electrical transport properties of the devices were measured down to $2\,\mathrm{K}$ using a closed-cycle cryostat, equipped with a high-precision low-current magnet power supply. In all cases the magnetic field was applied along the plane of the junctions. The magnetoresistance (MR) of the two device shown in Fig 3(a) and (c), measured at 10 K, are plotted in the Fig 3 (b) and (d), respectively. Clearly, the device with rougher interface exhibits a broader MR curve. Increased interface broadens the switching field distribution, leading to a delayed saturation \cite{zhao_roughness} of the magnetic films and, consequently, a broader MR curve. 

\begin{figure}[t]
    \centering
    \includegraphics[width=\columnwidth]{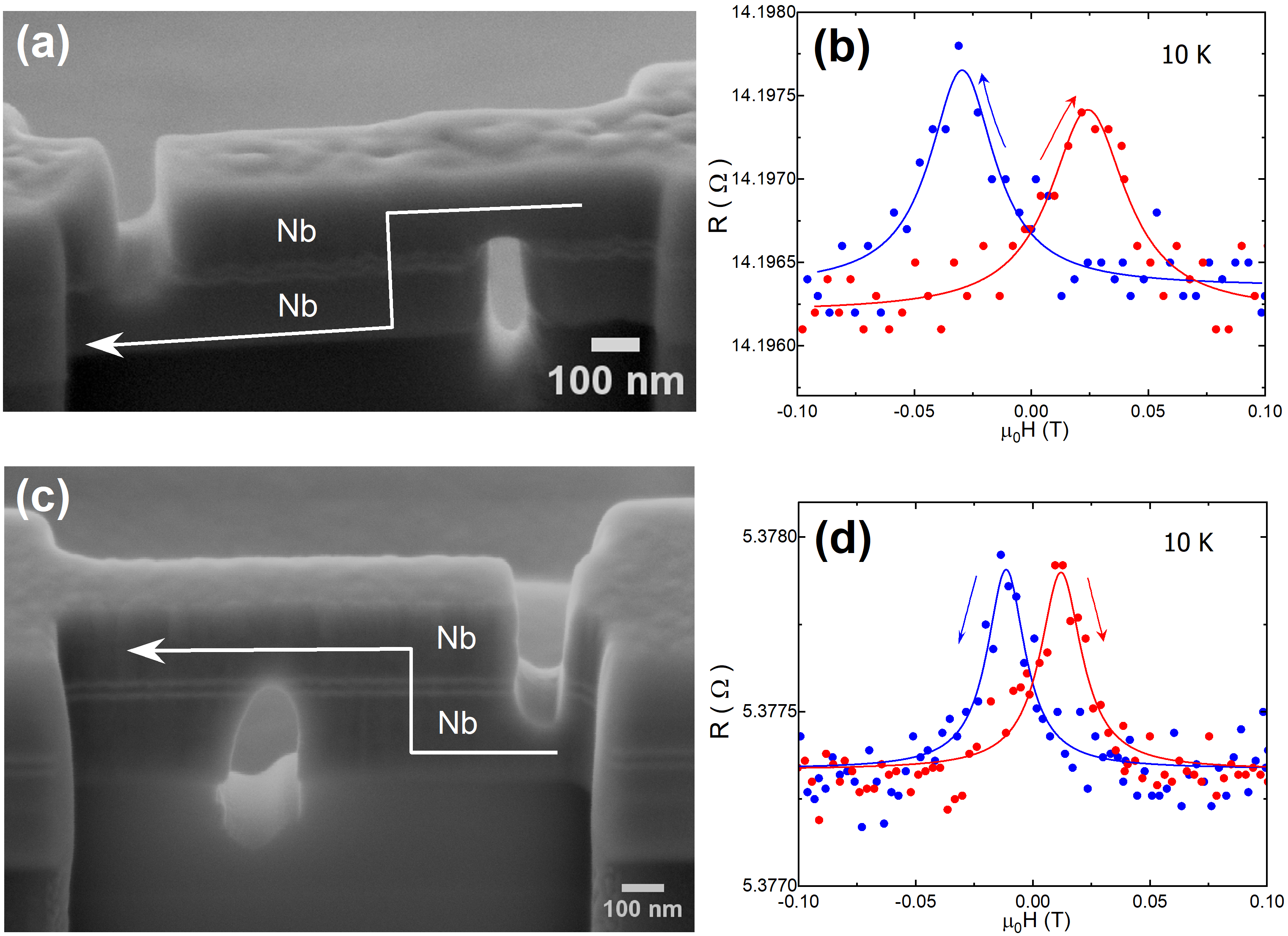} 
    \caption{Panels (a) and (b) show the scanning electron micrograph images of vertical nano devices fabricated from rough and smooth multilayers. Thickness of the multilayer was same in both cases as Nb(120 nm)/Pt(15 nm)/Ni(20 nm)/Pt(15 nm)/Nb(120 nm). The interface roughness in the Pt/Ni/Pt barrier is clearly discernible in panel(a), whereas a very smooth Pt/Ni/Pt barrier is observed in the panel(b). The arrow shows the current path, identifying the active area of the devices. The 10 K magnetoresistance, measured with in-plane magnetic field, are plotted in the panels (c) and (d) for the devices shown in panel (a) and (c), respectively. Rough device shows a wider MR curve corresponding to delayed magnetization rotation due to interface roughness. The solid lines are Lorentz fits to guide the eye.}
    \label{Fig.3}
\end{figure}
All devices fabricated from a series of Nb/Pt/Ni(20-35 nm)/Pt/Nb multilayers with smooth interfaces exhibited a completely flat R(T) curve down to the lowest available temperature well below the superconducting transition of the electrodes. High resolution electron microscope images of a few devices and their corresponding R(T) curves are show in the panels (a) and (c) of the supplementary Fig S1. Multiple transitions observed in the R(T) curves typically arise from the different widths of electrode tracks in FIB fabricated devices. These devices also exhibited the expected MR behavior at 10 K corresponding to the magnetization switching of the Ni layer, as shown in the panels (b) and (d) of the supplementary Fig S1. However, irrespective of the thickness of the Ni layer, the 2 K magneto-resistance (R(H)) curves were completely featureless in all devices with smooth interfaces. In comparison, the 2K MR of the devices with rough interface showed pronounced switching behavior, as shown in the Fig 4(a). More importantly, the direction of resistance switching was opposite during the forward and the reverse field sweeps, which is a clear signature of a spin valve behavior. In order to emphasize the magnitude and nature of the switching we have plotted the difference $\Delta R(H) $between the resistances of the reverse sweep branch and the forward branch of R(H) curves. Fig 4(b) compares the $\Delta R(H) $ curves, extracted from the Fig 4(a), for the smooth and rough devices. Clearly, the rough interface device has a significant resistance switching concurrent with the switching field of the Ni layer. A schematic representation of the process is shown in Fig 4(c). The non-equilibrium (in the current carrying state) spin magnetization (M$_S$) due to the equal-spin triplet component in the Pt layer adjacent to the superconducting Nb layer makes an effective valve-like arrangement with the Ni magnetization M$_{Ni}$. For a fixed direction of bias current, the direction of $M_s$ in Pt remains unchanged, irrespective of the direction of Ni magnetization. On the other hand, M$_{Ni}$ switches direction at coercive field while sweeping the magnetic field from positive saturation to negative saturation and vise-versa. Therefore, a higher resistance state due to anti-parallel arrangement of M$_S$ and M$_{Ni}$ rapidly switches to a lower resistance state when M$_S$ and M$_{Ni}$ become parallel. A change in the direction of bias current changes the direction of the induced triplet spin moment M$_S$. Therefore, upon reversal of the bias current the resistance switching direction also reverses, as seen in the Fig 4(d) for the 2 K data. This same general feature was reproducible in multiple devices across various thicknesses of Ni layer as shown in the supplementary Fig S2(c) and (d).

\section*{Discussion}

\begin{figure}[t]
    \centering
    \includegraphics[width=1.1\columnwidth]{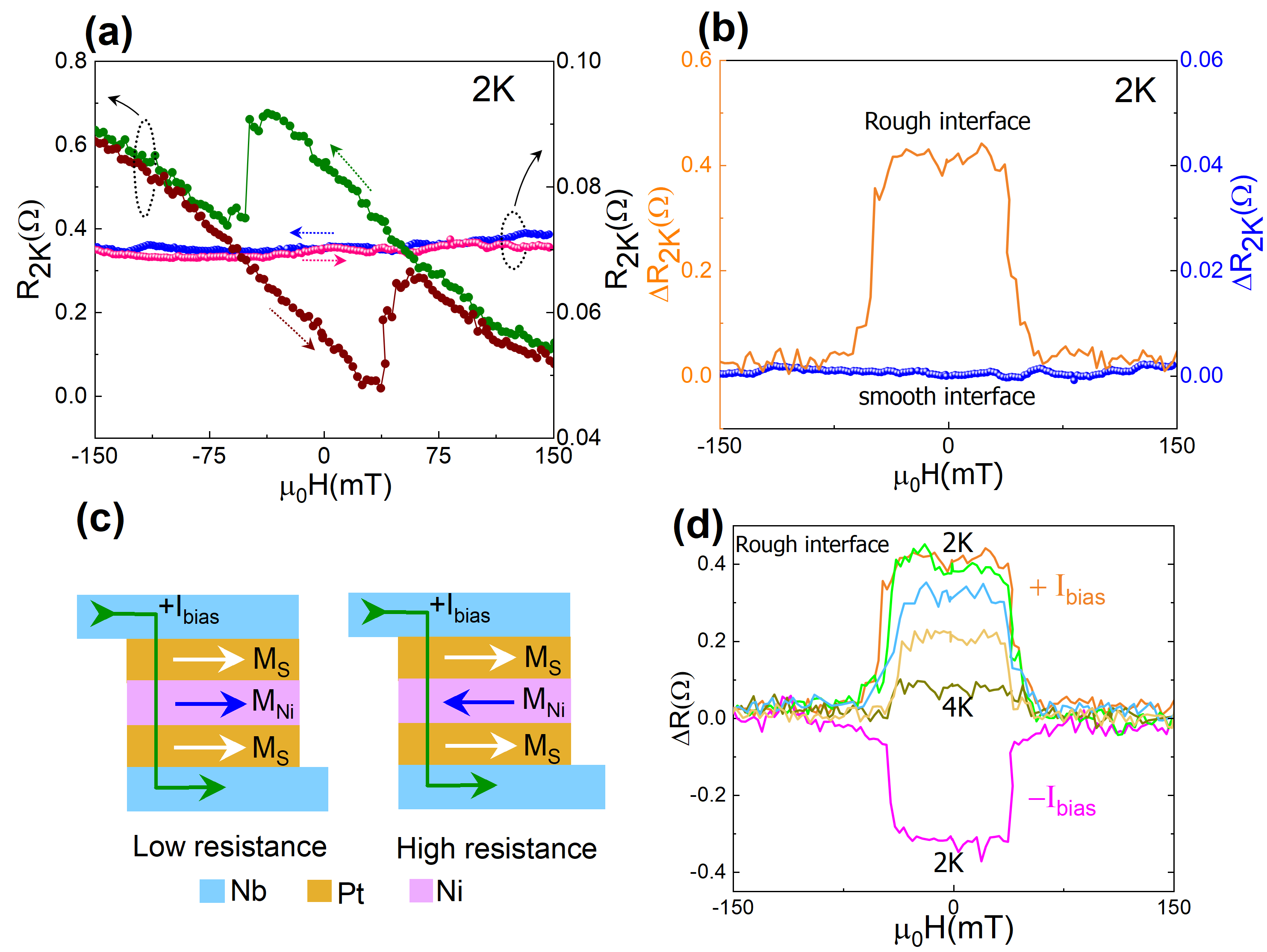}
    \caption{ (a) Comparison between the magnetoresistance R(H) curves of the smooth and rough devices, measured at 2 K. While the smooth device shows a featureless MR curve, the rough device shows pronounced switching effects in forward and reverse sweep data, though switching directions are different both cases. (b) The difference $\Delta R(H)$ between the R(H) curves are plotted, as extracted from the data in panel(a) to highlight the switching process. (c) A schematic view of the parallel and antiparallel configurations between the triplet spin moment M$_S$ in Pt and the magnetic moment M$_{Ni}$ in the Ni, which lead to the low and high resistance states. (d) $\Delta R(H)$ curves are plotted for the rough interface device at several temperatures between 2 K and 4 K, showing a strongly decreasing switching amplitudes. The effect of bias current reversal is also shown in the same figure. }
    \label{Fig.4}
\end{figure}

The presence of a valve-like behavior exclusively in devices with rough interface is consistent with the hypothesis that interfacial Rashba SOC enhances singlet-triplet conversion at the Nb/Pt interface, in the presence of a nearby magnetic exchange field. This observation is also consistent with our earlier results on relatively smoother interface Nb/Pt/Ni/Nb junctions, where the resistance difference between the parallel (P) and antiparallel (AP) configurations was approximately $1\,\mathrm{m\Omega}$. In comparison, the rough-interface junction exhibits a markedly enhanced valve effect, with a resistance difference of approximately $400\,m\Omega$ between the AP and P states \cite{SDEPtJJ}. Bergeret \textit{et al.}\cite{bergeret2014} predicted that, in a ferromagnetic Josephson junction, the interplay between the magnetic exchange field, Rashba and Dresselhaus spin orbit coupling can generate equal spin triplet correlations when the magnetization of the ferromagnetic layer is strictly in-plane. They also showed that in the presence of an out of plane magnetization only, the equal spin triplet component is not expected. The general condition for triplet generation given by Bergeret et al.\cite{bergeret2014} was modified by Satchell et.al \cite{satchell2019supercurrent} for the case of metallic systems where the contribution from Rashba spin-orbit coupling is nonzero ($\alpha \neq 0$), while the Dresselhaus contribution is zero ($\beta = 0$) \cite{satchell2019supercurrent}. They showed that for the generation of equal-spin triplet correlations in a superconductor/SOC/ferromagnet system, the exchange field must possess a canted magnetization. In combination with the Rashba SOC field, such a magnetization configuration enables the requisite spin rotation for generating equal-spin triplet correlations.Practically this can be achieved only if the magnetization of the Ferromagnetic layer has a combination of in-plane and out plane components. A similar condition for the generation of long-range triplet correlations is satisfied in our rough interface devices, because the Ni layer possesses an in-plane magnetic moment along with an roughness induced out of plane magnetic component. The wider coercive field ($H_c$) observed in the junction MR in Fig 3(b) and the higher saturation field observed in Fig 2(c) corroborates to this fact. Although a similar configuration was explored by Satchell \textit{et al.} using Co/Ni multilayers to achieve canted magnetization, no signature of triplet superconductivity was observed \cite{satchell2019supercurrent}. They attributed this to the micron-sized junctions, which could promote the formation of multiple magnetic domains within the ferromagnetic structure, thereby hindering the generation of triplet correlations. We noted that unlike in the case of smooth devices, a decreasing trend in resistance at the lowest temperatures (supplementary Fig S2(a) and (b)) was observed in the case of the rough interface devices, though no signatures of Josephson coupling was established. This delayed superconducting transition in the rough interface devices in supplementary Fig S2(a) and (b) could be a signature of the significant drainage of singlet cooper pairs due to triplet formation at the interface, as previously reported\cite{banerjee_triplet} by Banerjee et al. Establishment of a Josephson coupling across the Nb layers is, however, not a requirement for the functioning of the spin valve devices. What is required for spin-valve operation is the establishment of a long range triplet correlation in the Pt layer which, in presence of a bias current, can lead to a non-equilibrium spin moment. The possibility of a normal spin-hall effect contribution to the apparent spin moment in the Pt layer can be excludes because of two reasons. First, the spin valve effect rapidly diminishes with increasing temperature as shown in the Fig 4(d). This behavior is not expected for a quasiparticle dominated normal spin-Hall effect in Pt, since the quasiparticle population is expected to increase with increasing temperature. Second, since the conventional spin-Hall effect is a bulk phenomenon, the spin-valve type response would be expected in the smooth devices as well, irrespective of the interface roughness.

In conclusion, we have realized a spin-valve device combining the magnetic moment of a ferromagnetic layer with equal-spin triplet Cooper pair induced spin moment in a nonmagnetic Pt layer. Since these two components are independently controllable via magnetic field and bias current, these spin-valves can be operated through either parameters. In our devices the interface roughness of the multilayers, clearly discernible in FESEM images (in Fig 3(a), and supplementary Fig S2(a) and (b)), help the triplet generation process simultaneously in two different ways. First, as described in the Fig 2(b), the finite roughness at the Nb/Pt interface enables an interfacial Rashba SOC. Second, the same roughness also propagates into the Pt/Ni interface of the multilayer to result an out of place interfacial magnetization of Ni. Both effects are practically negligible in very smooth interface devices, which minimizes the chances of triplet generation, thereby excluding the possibility spin-valve response in the devices. These findings establish a novel route to directly integrate the polarization of the superconducting equal-spin triplet pair correlation into the conventional low temperature spintronics devices.

\section*{Methods}
\subsection*{Multilayer growth}
 Nb/Pt/Ni/Pt/Nb multilayer films were deposited on cleaned $5 \times 5$\,mm$^2$ Si/SiO$_2$ substrates using DC magnetron sputtering of high-purity Nb, Ni, Pt targets (99.998\%) at $\sim$1 Pa pressure. Prior to the deposition in an Ar ambient, a system base vacuum of the order of $2 \times 10^{-8}$\,mbar was achieved. In order to get a base vacuum with low H$_2$O and O$_2$ partial pressures the chamber was cycled through a 24 hours baking-cooling process, followed by Ti sublimation pumping. A residual gas analyzer (RGA) was used to monitor the partial pressure of water vapor and O$_2$ which are crucial for good superconducting properties of the Nb and a good interface transparency. The thickness of the various layers were calibrated in terms of exposure time to achieve consistent multilayer quality  \cite{Tapas}. In all cases the Nb and Pt layers were fixed at 160 nm and 15 nm, respectively. The thickness of the Ni layer were varied between 15 nm to 40 nm in various multilayers.

\subsection*{Device fabrication}
The nano-patterning of the junctions was carried out using focused ion beam (FIB) technique in a Zeiss Crossbeam 340 system. Electrode tracks were patterned using standard UV lithography. For nano-device fabrications a custom-designed stub with a 54\textdegree inclined face was used to mount the sample, allowing the required angular alignment between the sample surface and the FIB beam for milling from the top and side directions. In the first stage, an ion current of 100\,pA at 30\,kV was applied to narrow the 2$\mu$m track width to approximately 500\,nm. During the second stage, the current was reduced to 10\,pA at the same accelerating voltage to further decrease the width to about 200\,nm and to polish the sidewalls, thereby ensuring a clean and well-defined junction interface. These steps were performed with the ion beam oriented perpendicular to the sample surface. For the final stage, the sample was tilted such that the angle between the sample normal and the FIB beam was about 87\textdegree. A low beam current of 5\,pA at 30\,kV was then used to mill two narrow slots, each roughly 60\,nm wide, defining a nano-pillar with a rectangular cross-section.


\section*{Data Availability}
The data that support the findings of this study are available
within the Article and its Supplementary Information. Further information is available from the corresponding authors on reasonable request.

\bibliography{SSHE_spin_valve}

\begin{acknowledgments}
\vskip -0.5cm
Authors are grateful for financial support from NISER through
plan project RIN-4001 and Department of Science and Technology through project no:
ANRF/ARG/2025/007445/PS. 
\end{acknowledgments}

\section*{Author Contribution}
\vskip -0.5cm
DN performed the device fabrications and measurements. AS performed the magnetic characterization of films. DN, PKS and KS discussed and analyzed the data and wrote the manuscript. KS supervised the project.

\section*{Declaration} 
\vskip -0.5cm
The authors declare no competing interests

\setcounter{section}{0}
\setcounter{figure}{0}
\setcounter{table}{0}
\setcounter{equation}{0}

\renewcommand{\thesection}{\arabic{section}}
\renewcommand{\thefigure}{S\arabic{figure}}
\renewcommand{\thetable}{S\arabic{table}}
\renewcommand{\theequation}{S\arabic{equation}}
\clearpage
\onecolumngrid

\begin{center}
{\Large\bfseries Supplementary data}\\[1em]

{\LARGE\bfseries
Interface engineering of spin triplet Cooper pairs for spin-valve implementation
}\\[1em]

Debashree Nayak$^{1,2}$, Abhisek Sahoo$^{1,2}$, Pratap Kumar Sahoo$^{1,2}$, Kartik Senapati$^{1,2}$,
\\[0.5em]

$^{1}$ School of Physical Sciences, National Institute of Science Education and Research (NISER) Bhubaneswar, Jatni, Odisha-752050, India\\
$^{2}$ Homi Bhabha National Institute, Training School Complex, Anushakti Nagar, Mumbai-400094, India
\end{center}

\vspace{2em}

\section*{Additional devices with smooth interface}

\begin{figure}[!h]
    \centering
    \includegraphics[width=0.9\textwidth]{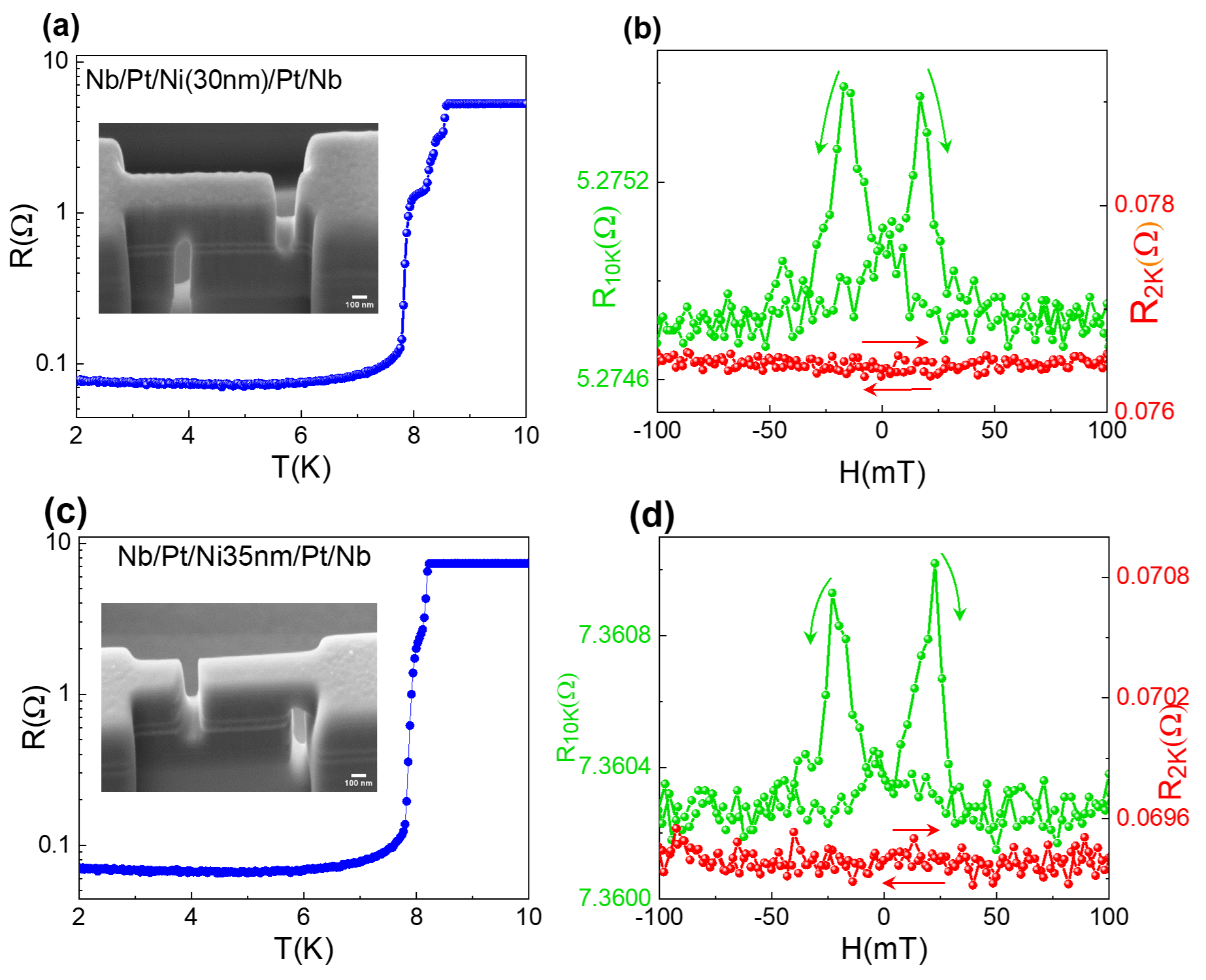}
    \caption{(a) and (c) show the SEM images and the respective R(T) curves of Nb/Pt/Ni/Pt/Nb vertical nano device with 30 nm and 35 nm thick Ni layers, respectively. The three layer smooth Pt/Ni/Pt barriers are clearly visible in these images. Panels (b) and (d) plot the 2 K and 10 K MR curves of the devices shown in (a) and (c), respectively. No switching feature was visible in the 2 K data.}
\end{figure}

\section*{Additional devices with rough interface}

\begin{figure}[!h]
    \centering
    \includegraphics[width=0.9\textwidth]{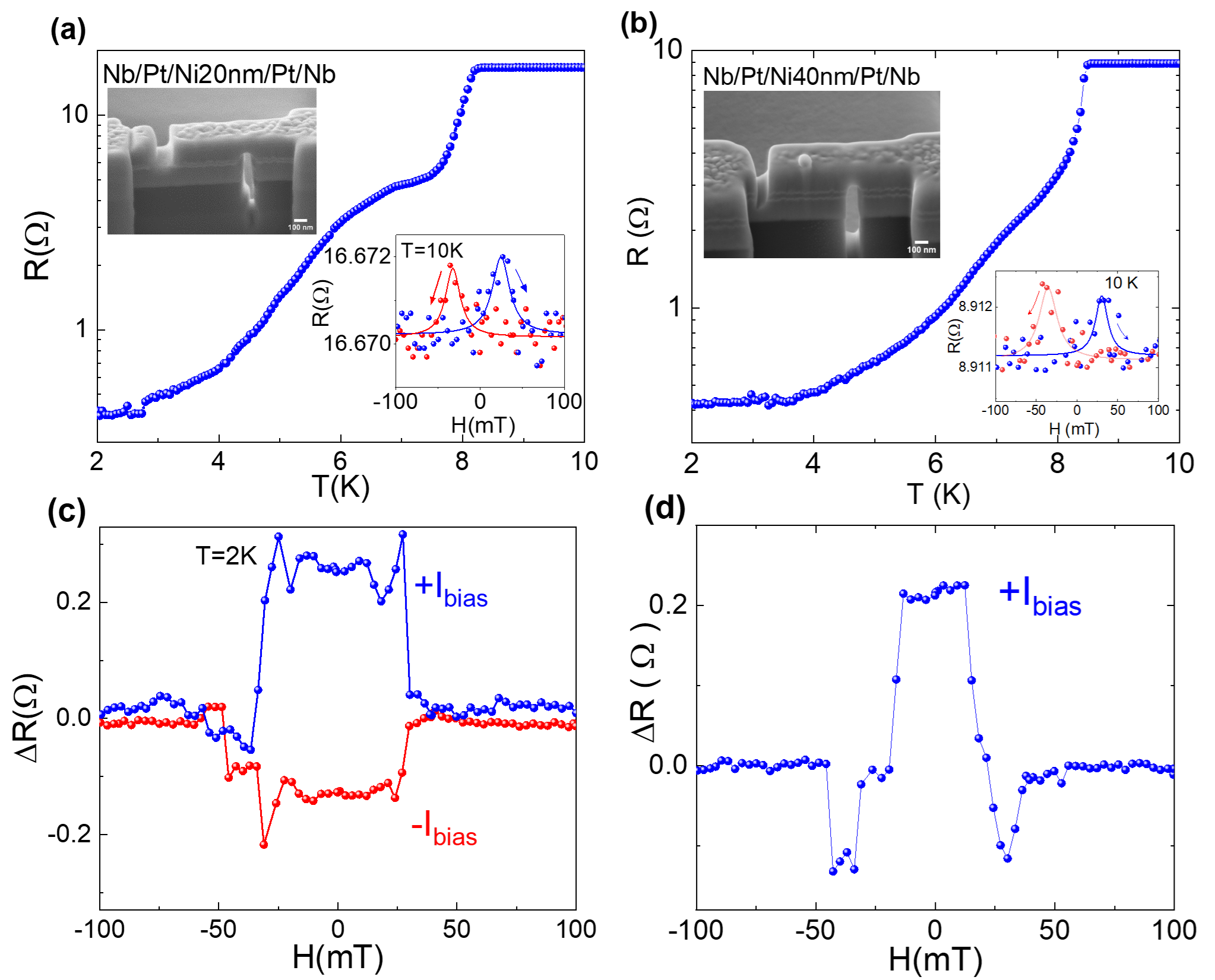}
    \caption{(a) and (b) show the SEM images and the respective R(T) curves of Nb/Pt/Ni/Pt/Nb vertical nano device with 20 nm and 40 nm thick Ni layers, respectively. The three layer Pt/Ni/Pt barriers with rough interface are clearly visible in these images. Insets show the usual MR response of the devices at 10 K. Panels (c) and (d) plot the $\Delta R$(H) curves clearly demonstrating the valve-like response upon magnetic field induced switching of Ni moment for positive and negative bias currents. }
\end{figure}

\end{document}